\documentclass[a4paper,notitlepage,final]{article}
\usepackage{amsmath}
\usepackage{amssymb}
\usepackage{xspace}
\usepackage[numbers]{natbib}

\usepackage{tikz}
\usetikzlibrary{positioning}
\usepackage{graphicx}

\usepackage{diagbox}
\usepackage[boldmath]{numprint} %Defines new column descriptors n and N
\usepackage{colortbl}
\usepackage[pdftex,pdfstartview=FitH,linkcolor=blue,linkbordercolor=blue,colorlinks=false,pdfpagemode=UseOutlines]{hyperref} % so that it opens page=bounding box
\usepackage[all]{hypcap} % to link to the top of Figure or Table

\usepackage[ansinew]{inputenc}
\usepackage[T1]{fontenc}
\newcommand\comb[1]{\texttt{#1}}
\newcommand\eg{e.g.,\xspace}
\newcommand\ie{i.e.,\xspace}
\newcommand\free{\textit{free}\xspace}
\newcommand\zero{\textit{zero}\xspace}
\newcommand\ccell[1]{\multicolumn{1}{c|}{#1}}
\newcommand{\myb}[1]{{\npboldmath} #1} % to put result in bold using numprint

\begin{document}

\npstyleenglish

\hypersetup{%
  pdfauthor={Geoffroy Ville},
  pdftitle={An Optimal Mastermind (4,7) Strategy and More Results in the Expected Case},
  pdfsubject={Optimal Mastermind Solution},
  pdfkeywords={Mastermind,optimal,solve,expected case,symmetry,case equivalence,signature,free color,zero color,representative,upper-bound,lower-bound}}

\title{An Optimal Mastermind (4,7) Strategy and More Results in the
  Expected Case}
\author{Geoffroy \textsc{Ville}\footnote{< gville. mastermind at gmail. com >}}
%\email{gville.mastermind at gmail.com}
\date{March 2013}

\maketitle

\begin{abstract}
  This paper presents an optimal strategy for solving the 4 peg-7 color
  Mastermind MM(4,7) in the expected case (4.676) along with optimal
  strategies or upper bounds for other values. The program developed is
  using a depth-first branch and bound algorithm relying on tight upper
  bound, dynamic lower bound evaluation and guess equivalence to prune
  symmetric tree branches.
\end{abstract}

\section{Introduction}
Mastermind is a code-breaking game for two players. This peg game was
invented in 1970 by Mordecai Meirowitz. One of the players -- the
\textit{code-maker} -- chooses a secret code of four pegs of six possible
repeatable colors, thus one code among 1296 possibilities. The other player
-- the \textit{code-breaker} -- tries to break the code by making guesses,
\ie submitting one code. The code-maker answers using four pegs of two
colors: a black peg means that a guess peg matches both color and position
of a code peg, whereas a white peg means that a guess peg matches the color
but not the position of a code peg. An answer containing less than four
pegs indicates that one or more guess pegs are not part of the code. The
answer is global so the code-breaker does not know which black/white peg
corresponds to which guess peg. The code-breaker has 8 to 12 guesses
(depending of the board) to find the correct code.

The classic board game MM(4,6) can also be played with a virtual seventh
color by leaving one or several holes in the code, MM(4,7). It can be
generalized to any number of pegs or colors, MM(p,c).

Much research has been done on this game and its
variants. \citet{Knuth1976} proposed a strategy that requires 4.478
guesses, on average, to find the code (expected case) while always finding
the code in a maximum of 5 guesses (worst case). While 5 guesses is the
optimal (minimal worst case), several authors proposed different other
one-step-ahead heuristic algorithms to reduce the average number of cases
until \citet{Koyama1993} found the optimal strategy for the expected case
to be 4.340 using a depth-first backtracking algorithm. This optimal
strategy requires 6 guesses. Some authors also solved theoretically a few
expected cases (\citet{Chen2004}, \citet{Goddard2004}) or worst cases
(\citet{Jager2009}).

Higher cases were soon tackled through different approaches like genetic
algorithms (\citet{Merelo2006}) and upper bounds of the expected
case were found.

The game of Bulls and Cows, which may date back a century or more and
does not allow color repetition, is also worthy of mention as some
Mastermind research started on this problem (\citet{ChenN2004}).

Another variant called Static Mastermind is also popular. In this game, all
the guesses are made at once. Then, given all the answers, the code can be
deduced with one additional guess. Minimizing this number of guesses also
led to much research (\eg \citet{Goddard2003} or \citet{CutTheKnot}).

To pursue the optimal strategy track in the expected case and to take the
list of results published by \citet{Goddard2004} a step further, a
depth-first branch and bound algorithm with a tight upper bound at the
start, a dynamic lower bound evaluation during the resolution, and
detecting guess-equivalence at each step using the symmetries of the colors
that were played was developed.

Section~2 defines mathematical notations and ways of presenting the
problem. Section~3 reviews in more details past work on both MM(4,6) and
MM(5,8) and presents or reproduces results of detailed heuristic
models. Section~4 describes theoretical established results. Section~5
explains in detail ways to reduce computing time of the program used to
find MM(4,7). Section~6 presents results on the expected value. The
conclusion makes suggestion for further research.

\section{Definitions}

\subsection{Mathematical notations}
Throughout this article, the following mathematical notations will be used:
\begin{flalign}
c         &= \text{\# colors}                                     &                   \\
p         &= \text{\# pegs}                                       &                   \\
N_{p,c}   &= \text{\# possible codes with $p$ pegs and $c$ colors} &= c^p              \\
G_p       &= \text{\# grades  with $p$ pegs (and $c$ colors)}     &= \frac{p(p+3)}{2} \\
E(p,c)    &= \text{best average in the expected case for MM(p,c)} &                   \\
W(p,c)    &= \text{best worst case for MM(p,c)}                   &                   \\
MM^*(p,c) &= \text{MM(p,c) with possible guesses only}            & 
\end{flalign}

While Equation 3 is obvious, Equation 4 requires more explanation. It can
be obtained two ways.

The first one is inferred from \autoref{tab-eval4trous}. All grading
possibilities are on the upper left part of the table. Moreover, on the
diagonal, all gradings are possible but $(p-1,1)$ for obvious reasons. Thus,
\[G_p=\frac{\overbrace{(p+1)\times(p+1)}^{whole~table}-\overbrace{(p+1)}^{diagonal}}{2}+p = \frac{(p+1)p+2p}{2}=\frac{p(p+3)}{2}\].

The second way comes from the fact that $G_p$ can also be seen as the
number of ways to solve $b+w+z=p$ ($z$ integer), minus the impossible
solution of $(b,w)=(p-1,1)$. Therefore,
\[G_p= \binom{p+3-1}{3-1}-1=\frac{(p+2)!}{2!p!}-1=\frac{(p+1)(p+2)-2}{2}=\frac{p^2+3p}{2}=\frac{p(p+3)}{2}\].

In the MM(4,7) case, $N_{4,7}=2401$ and $G_4=14$.

\begin{table}
  \centering
  \caption{The 14 possible grades in MM(4,7)}
  \label{tab-eval4trous}
  \begin{tabular}{|c|ccccc|}\hline
    \diagbox{b}{w} & 0     & 1     & 2     & 3     & 4    \\ \hline
    0              & (0,0) & (0,1) & (0,2) & (0,3) &(0,4) \\
    1              & (1,0) & (1,1) & (1,2) & (1,3) & x    \\
    2              & (2,0) & (2,1) & (2,2) & x     & x    \\
    3              & (3,0) & x     & x     & x     & x    \\
    4              & (4,0) & x     & x     & x     & x    \\ \hline
  \end{tabular}
\end{table}
% \begin{table}
%   \centering
%   \caption{The 14 possible grades in MM(4,7)}
%   \label{tab-eval4trous}
%   \begin{tabular}{|c|c|c|c|c|c|}\hline
%     \diagbox{b}{w} & 0     & 1     & 2     & 3     & 4     \\ \hline
%     0              & (0,0) & (0,1) & (0,2) & (0,3) & (0,4) \\ \hline
%     1              & (1,0) & (1,1) & (1,2) & (1,3) &       \\ \cline{1-5}
%     2              & (2,0) & (2,1) & (2,2) & \multicolumn{2}{c|}{} \\ \cline{1-4}
%     3              & (3,0) & X     & \multicolumn{3}{c|}{}  \\ \cline{1-3}
%     4              & (4,0) & \multicolumn{4}{c|}{}  \\ \hline
%   \end{tabular}
% \end{table}

\subsection{Summing by colors}
\label{sec:sumcolors}

The total number of codes can also be thought of as the sum of all the
number of possible combinations of $i$ colors.

Let $C_i$ be the number of possibles codes of $p$ pegs of exactly $i$
colors chosen among $c$. $C_i$ is the way of choosing $i$ colors among $c$
multiplied by the number of ways of setting these $i$ colors in a code of
length $p$, denoted $Z_{pi}$. As there are no more than $p$ colors and $c$
possibilities,
\[\forall (p,c), \, N_{p,c}=c^p=\sum_{i=1}^{min(c,p)}{C_i}=\sum_{i=1}^{min(c,p)}{\binom{c}{i}Z_{pi}}\].

where $Z_{pi}$ is the sum of all possible distributions of the $i$ colors and given by the following formula:
\[Z_{pi}=\sum_{n_1+n_2+ \cdots +n_i=p}\frac{(n_1+n_2+ \cdots +n_i)!}{n_1!n_2!\cdots
  n_i!}=\sum_{n_1+n_2+ \cdots +n_i=p}\frac{p!}{n_1!n_2!\cdots n_i!}\]

$Z_{p1}=1$ because there is only one way of placing $p$ pegs of the same
color and, $\forall p \leq c, Z_{pp}=p!$ because it is the number of ways
of coding $p$ pegs of at least $p$ colors.

For example, in the MM(4,7) game,
\begin{equation*}
\begin{split}
N_{4,7} & = \binom{7}{1}1+\binom{7}{2}(\frac{4!}{3!1!}+\frac{4!}{2!2!}+\frac{4!}{1!3!})+\binom{7}{3}(3\times\frac{4!}{1!2!1!})+\binom{7}{4}4! \\
%       & = 7\times1+21\times14+35\times36+35\times24 \\
       & = 7+294+1260+840                             \\
       & = 2401                                      
\end{split}
\end{equation*}

%The number of distributions $k$ is
%$\binom{p+k-1}{k-1}=\frac{(p+k-1)!}{p!(k-1)!}$.

\subsection{Modeling a solution as a tree}

% Define styles for bags and leafs
%\tikzstyle{comb} = [circle, text width=4em, text centered, draw=black]
\tikzstyle{comb}=[circle,text centered,draw]
\tikzstyle{found}=[circle,inner sep=0pt]
\tikzstyle{answer}=[pos=0.5,fill=white,inner sep=2pt]

\begin{figure}
\begin{centering}
\begin{tikzpicture}[level/.style={sibling distance=30mm/#1}]
\node [comb] (a) {112}
  child {node [comb] (b) {221}
    child [grow=right] {[fill] circle (2pt)} 
    edge from parent node[answer] {02}
  } 
  child {node [comb] (d) {222} 
    child [grow=right,level distance = 8mm] {[fill] circle (2pt)}
    edge from parent node[answer] {10}
  } 
  child {node [comb] (f) {121}
    child {node [comb] (g) {211}
      child [grow=right,level distance = 8mm] {[fill] circle (2pt)}
      edge from parent node[answer] {12}
    } 
    child [grow=right,level distance = 8mm] {[fill] circle (2pt)}
    edge from parent node[answer] {12}
  }
  child {node [comb] (j) {122}
    child {node [comb] (k) {111}
      child [grow=right,level distance = 8mm] {[fill] circle (2pt)}
      edge from parent node[answer] {10}
    }
    child {node [comb] (m) {212}
      child [grow=right,level distance = 8mm] {[fill] circle (2pt)}
      edge from parent node[answer] {12}
    }
    child [grow=right,level distance = 8mm] {[fill] circle (2pt)}
    edge from parent node[answer] {20}
  }
  child [grow=right,level distance = 8mm] {[fill] circle (2pt)}
;
\node (l3) [right = of m] {= 3}
child [grow=up] {node (l2) {= 4} edge from parent[draw=none]
  child [grow=up] {node (l1) {= 1} edge from parent[draw=none]}
};

%\path (p) -- (l1) node [midway] {=};
%\path (o) -- (l2) node [midway] {=};
%\path (m) -- (l3) node [midway] {=};

\end{tikzpicture}
\end{centering}
\caption{A MM(3,2) optimal tree and the number of codes found at each
  level}
\label{fig:treeopt32}
\end{figure}
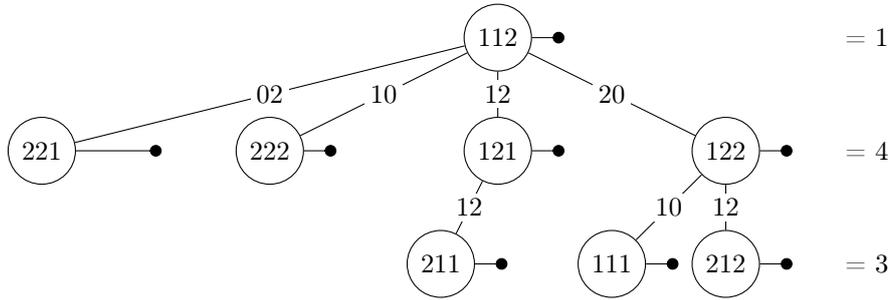

Let's picture a tree representation of a solution
(\autoref{fig:treeopt32}). Each node is a guess and each branch an
answer. A $(p,0)$ branch leads to an end node, also called a leaf node. Any
given solution tree has two attributes, its depth\footnote{A tree with a
  single node has a depth of 0.} $D$ and its external path
length\footnote{The external path length is the sum of the path lengths
  to each leaf node.} $L$.

To compare different strategies, a uniform distribution of codes is
assumed. Consequently, the two complementary ways of measuring a
strategy performance are the maximum number of guesses needed in the
\textit{worst case} $W$, and the average number of guesses in the
\textit{expected case} $E$. They can be expressed in terms of $D$ and
$L$ as follows:
\begin{align}
W &= D+1 \\
E &= \frac{L}{N}
\end{align}

In \autoref{fig:treeopt32}, $D=2$ and
$L=1\times1+2\times4+3\times3=18$, thus $W=3$ and
$E=\frac{18}{8}=2.250$.  More generally, if $f_i$ is the number of
codes \textit{found} at guess $i$, then $N=\sum_{i=1}^{w}{f_i}$ and
$L=\sum_{i=1}^{w}i{f_i}$.

Minimizing $W$ and minimizing $E$ are two different goals. For
example, the optimal value of $E$ for MM(4,6) is
$E(4,6)=5625/1296=4.340$, with 6 moves in the worst
case\citep{Koyama1993}. But \citet{Knuth1976} showed that one could
always find the answer in $5<6$ questions ($W(4,6)=5$ in fact) with an
increase in the number of moves.

\begin{figure}
  \centering
  \includegraphics[width=\textwidth]{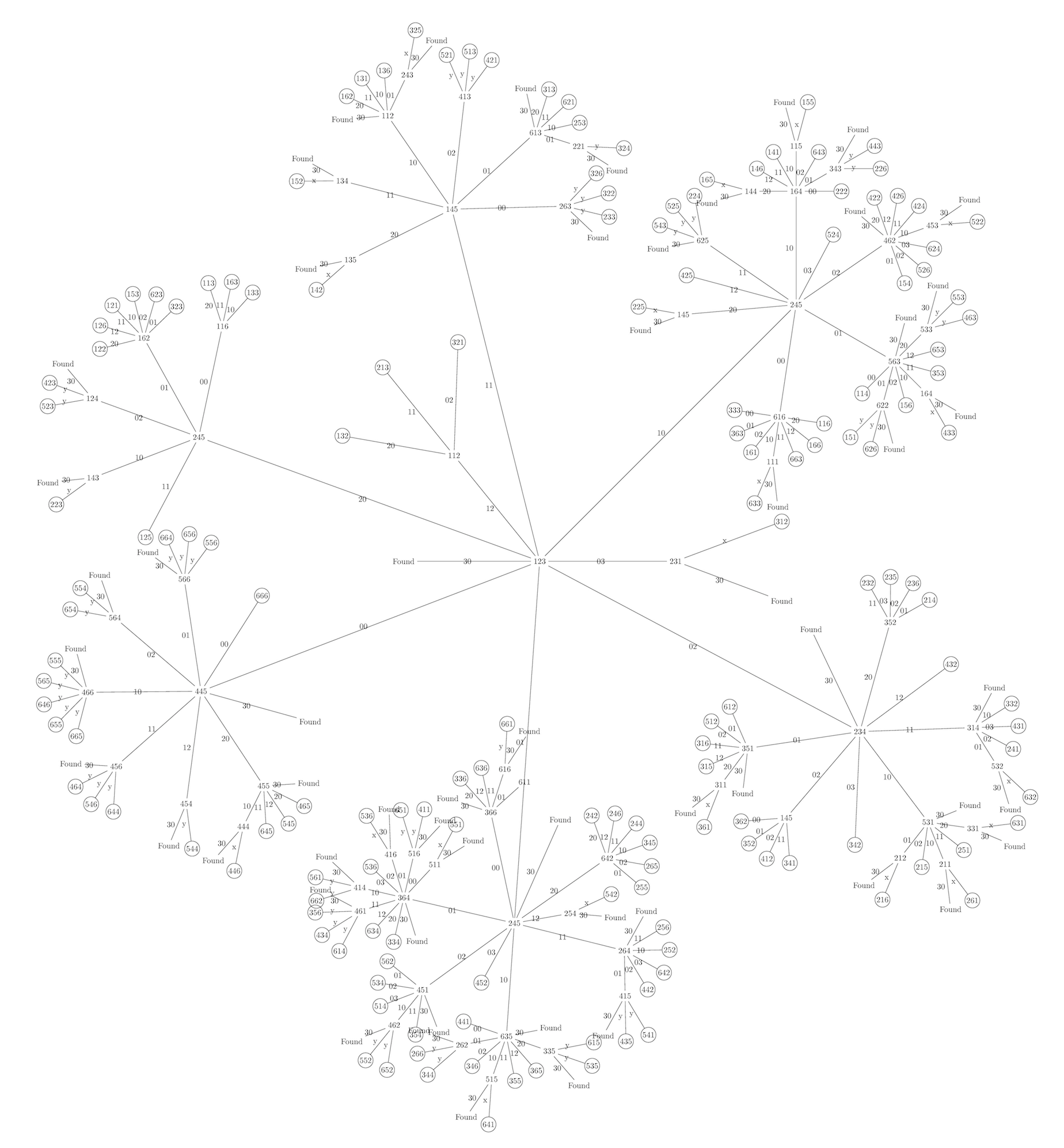}
  \caption{Full `flower' of solutions for MM(3,6)}
  \label{fig:mm36}
\end{figure}

Finally, as both an artistic representation and an illustration of the
unbalanced nature of a solution tree, \autoref{fig:mm36} depicts an
optimal solution of MM(3,6) in the form of a circular tree or flower.

\subsection{A peculiar grading function}
Let $g$ be the grading function. $g$ is commutative, \ie given 2 codes $a$
and $b$, $g(a,b)=g(b,a)$. Therefore, if a guess $a$ is graded (2,0) for
example, the solution lies among the codes graded (2,0) with $a$.

All Mastermind algorithms are based on this fundamental property.

\section{Solving Mastermind through heuristics}

\subsection{Guessing only possible codes is not optimal}
\label{sec:nonpossible}
Knuth was the first to show that using guesses that are not valid codes
when played could reduce the solution tree depth.

This remarkable result is not intuitive and is illustrated by
\autoref{tab:nonpossiblecode}. It shows a case in which 6 codes are left to
be found, and guessing a seventh discriminates all the codes in one
question ($L=1\times0+2\times6=12$) whereas playing any of the 6 codes
requires at least two ($L=1\times1+2\times3+3\times2=13$ in the case of 4
different answers).

For example, when solving MM(4,6) using Knuth's algorithm but with
\textit{possible} guesses only, namely MM*(4,6)\footnote{We will see in
  \autoref{sec:upperbound} how solving a possible case can still be
  useful.}, the result goes up to $5828/1296=4.497$; after 2 guesses only, 7
(1+6) codes are found while the original algorithm reveals 14 (1+13) codes
after two guesses.

Therefore and unfortunately, the number of codes to try at each step
is not the monotonously decreasing set that one would hope for, it
remains constant minus the previous guesses. Solving the problem
requires memory and processor time. Some well-known heuristics have
been developed to this end.

\begin{table}
  \centering
  \caption{How a non possible code can segregate a set of possible ones}
  \label{tab:nonpossiblecode}
  \begin{tabular}{c|cccccc|c} \hline
 c & \comb{121} & \comb{122} & \comb{126} & \comb{153} & \comb{323} & \comb{623} & \textbf{162} \\ \hline
 \comb{121} & (3,0) & (2,0) & (2,0) & (1,0) & (1,0) & (1,0) & (1,1) \\ \hline
 \comb{122} & (2,0) & (3,0) & (2,0) & (1,0) & (1,0) & (1,0) & (2,0) \\ \hline
 \comb{126} & (2,0) & (2,0) & (3,0) & (1,0) & (1,0) & (1,1) & (1,2) \\ \hline
 \comb{153} & (1,0) & (1,0) & (1,0) & (3,0) & (1,0) & (1,0) & (1,0) \\ \hline
 \comb{323} & (1,0) & (1,0) & (1,0) & (1,0) & (3,0) & (2,0) & (0,1) \\ \hline
 \comb{623} & (1,0) & (1,0) & (1,1) & (1,0) & (2,0) & (3,0) & (0,2) \\ \hline
 Diff. grades & 3    & 3     & 4     & 2     & 3     & 4     & \textbf{6} \\ \hline
  \end{tabular}
\end{table}
% From Kooi2005, avoid copyright problems
% \begin{table}
%   \centering
%   \caption{How a non possible code can segregate a set of possible ones}
%   \label{tab:nonpossiblecode}
%   \begin{tabular}{c|cccccc|c} \hline
%  c & \comb{1211} & \comb{1212} & \comb{1216} & \comb{1245} & \comb{1515} & \comb{1615} & \textbf{1261} \\ \hline
%  \comb{1211} & (4,0) & (3,0) & (3,0) & (2,0) & (2,0) & (2,0) & (3,0) \\ \hline
%  \comb{1212} & (3,0) & (4,0) & (3,0) & (2,0) & (2,0) & (2,0) & (2,1) \\ \hline
%  \comb{1216} & (3,0) & (3,0) & (4,0) & (2,0) & (2,0) & (2,1) & (2,2) \\ \hline
%  \comb{1245} & (2,0) & (2,0) & (2,0) & (4,0) & (2,0) & (2,0) & (2,0) \\ \hline
%  \comb{1515} & (2,0) & (2,0) & (2,0) & (2,0) & (4,0) & (3,0) & (1,1) \\ \hline
%  \comb{1615} & (2,0) & (2,0) & (2,1) & (2,0) & (3,0) & (4,0) & (1,2) \\ \hline
%  Diff. grades & 3    & 3     & 4     & 2     & 3     & 4     & \textbf{6} \\ \hline
%   \end{tabular}
% \end{table}

\subsection{Simple consistency algorithm}
The first algorithm one might think of programming is extremely simple yet
efficient. It starts with the first code \comb{1111} then:
\begin{enumerate}
\item submits a guess and stores the result,
\item chooses, in lexical order, the next guess compatible with the stored
  results, and
\item stops when $(b,w)=(p,0)$.
\end{enumerate}

On the one hand, its main advantages are simplicity and very low
computation and memory cost. Only previous guesses and grades are
stored. Each new code is created along the way and graded against the codes
in history. There is no need to store the set of possible solutions at each
step. On the other hand, since it only uses possible codes and does not try
to aim for any optimization, its value is very high. When starting with
\comb{1111}, the expected case is $7471/1296=5.765$, with 9 guesses in the
worst case. \citet{Shapiro1983}\footnote{I did not have access to that
  paper.} described this strategy.

Unlike the algorithms described in \autoref{sec:heuristics}, it chooses the
next possible solution solely on the lexical order basis. Thus, if the
first guess makes the program jump faster toward a suitable reduced range
of possible codes, the result improves. \citet{Swaszek2000}\footnote{I did
  not have access to that article either and quote it through what other
  articles describe.}  tried choosing the guesses randomly and found an
expected case of 4.638.

A non random approach can also improve the result. The first possible code
in lexical order is still chosen at each step. If
\comb{1122}\footnote{\comb{1122} is among the `best' first guess of
  \autoref{tab:comparealgo46}.} is forced as the first guess, the expected
case reduces to $6508/1296=5.021$ with 8 guesses. If the first guess is
\comb{3456}, giving information about the colors that are usually explored
at the very end, the result becomes $6045/1296=4.664$ and 7 guesses.

As shown in the next section, there is no best first code: depending on
which algorithm is used, one code suits it best.

\subsection{One-step-ahead heuristics}
\label{sec:heuristics}
Other algorithms look one step ahead and choose the next guess
accordingly. All these algorithms sort, for all guesses, the remaining
possible solutions by grade into subsets. They differ by the way, based on
these subsets, they select their next guess. When several guesses are
eligible, the final choice is made by selecting the possible ones, if any,
among them and ultimately choosing the first in lexical order.

Limitations are twofold. Obviously, starting with the best first-step guess
only optimizes the next step, while another might have optimized the next
two steps and be globally better. Using another sorting policy than lexical
order among the eligible guesses at each step has an impact on the results
which shows that guesses are not all equivalent (impact on the following
steps).

\citet{Knuth1976} showed that a guess that would minimize the maximum
number of codes in each of $k$ subsets (\textit{Max.~size} algorithm,
minimize $max(n_i)_{i=1..k}$) has an average of $5801/1296=4.476$. The best
first code (lexical order, one step ahead) for this algorithm is
\comb{1122}.

\begin{table}
  \centering
  \caption{One-step-ahead algorithm results for MM(4,6) vs. the optimal}
  \label{tab:comparealgo46}
  \begin{tabular}{|l|c|n{4}{0}|n{1}{3}|c|} \hline
Algorithm    & First Guess & \ccell{L}  & \ccell{E}   & W          \\ \hline
Max. size    & \comb{1122} & 5801       & 4.476       & \textbf{5} \\ \hline
Expect. size & \comb{1123} & 5696       & 4.395       & 6          \\ \hline
Entropy      & \comb{1234} & 5723       & 4.416       & 6          \\ \hline
Most parts   & \comb{1123} & 5668       & 4.373       & 6          \\ \hline
Optimal      & \comb{1123} & \myb{5625} & \myb{4.340} & 6          \\ \hline
  \end{tabular}
\end{table}

The next guess can also be selected by minimizing the expected size of the
$k$ subsets (\citet{Irving}\footnote{Irving used this strategy for the
  first two guesses and did an exhaustive search after. The pure strategy
  was used here for all the steps.}, minimize
$\sum_{1}^{k}{\frac{n_i^2}{N}}$), maximizing their entropy
(\citet{Neuwirth1982}\footnote{I did not have access to his article.The
  pure strategy was kept.}, maximize $\sum_{1}^{k}{-p_iLog_2(p_i)}$ with
$p_i=n_i/N$) or maximizing their number (\citet{Kooi2005}, maximize $k$).

\autoref{tab:comparealgo46} summarizes the results for
MM(4,6). \textit{Most parts} gives the lowest score in expected value
($5668/1296=4.373$) while \textit{Max.~size} gives the lowest number of
guesses in the worst case (5). Recreating this table, the same results as
\cite{Kooi2005} were found except for \textit{Entropy} (5723 versus
5722\footnote{When running the program taking the last code in lexical
  order, 5722 is found.}).

The results of these algorithms on the MM(4,7) case are presented in
\autoref{tab:comparealgo47}. This time, \textit{Entropy} performs
better. As for MM(4,6), different results are obtained when starting
with different first guesses or choosing an order other than lexical.

\begin{table}
  \centering
  \caption{One-step-ahead algorithm results for MM(4,7)}
  \label{tab:comparealgo47}
  \begin{tabular}{|l|c|n{5}{0}|n{1}{3}|c|} \hline
    Algorithm    & First Guess & \ccell{L}   & \ccell{E}   & W \\ \hline
    Consistency  & \comb{4567} & 12265       & 5.108       & 8 \\ \hline
    Max. size    & \comb{1234} & 11613       & 4.837       & 6 \\ \hline
    Expect. size & \comb{1234} & 11409       & 4.752       & 6 \\ \hline
    Entropy      & \comb{1234} & \myb{11382} & \myb{4.740} & 6 \\ \hline
    Most parts   & \comb{1123} & 11388       & 4.743       & 6 \\ \hline
  \end{tabular}
\end{table}

The results of these algorithms for MM(5,8) were compared with the results
of \citet{Heeffer2008}. The standard deviation algorithm introduced in the
paper was not implemented because of its poor performance. The results
found, presented in \autoref{tab:comparealgo58}, are very different. The
result of \textit{Consistency} cannot be compared with the random runs
performed by the authors. But for \textit{Max.~size}, 5.614 was found
versus 5.670, for \textit{Expect.~size} 5.502 versus 5.601 and for
\textit{Entropy} 5.489 versus 5.583. For \textit{Most parts}, \comb{11223}
was used as the first guess (first in lexical order) rather than
\comb{11234} reported in the article. Still, the difference is significant
(5.549 versus 5.693).  The source of the discrepancies was not
found.\footnote{All the MM(4,6) results are quoted in the article and
  supposedly reproduced by the authors. Furthermore, for MM(5,8), the table
  of partitions and choice for the first guess are the same. The authors
  could not be reached.} Even though these differences, the algorithms
perform in the same order: \textit{Entropy} is the winner, followed by
\textit{Expect.~size}, \textit{Most parts}, \textit{Max.~size} and finally
\textit{Consistency}. In terms of worst case, \textit{Entropy} has one code
that requires 8 guesses. There should be a way to find 7 guesses by
increasing the expected average.

\begin{table}
  \centering
  \caption{One-step-ahead algorithm results for MM(5,8)}
  \label{tab:comparealgo58}
  \begin{tabular}{|l|c|n{6}{0}|n{1}{3}|n{2}{0}|} \hline
    Algorithm    & First Guess  & \ccell{L} & \ccell{E}   & \ccell{W} \\ \hline
    Consistency  & \comb{45678} & 195633    & 5.970       & 10        \\ \hline
    Max. size    & \comb{11234} & 183966    & 5.614       & \myb{7}   \\ \hline
    Expect. size & \comb{11234} & 180287    & 5.502       & \myb{7}   \\ \hline
    Entropy      & \comb{11234} & 179879    & \myb{5.489} & 8         \\ \hline
    Most parts   & \comb{11223} & 181834    & 5.549       & 9         \\ \hline
  \end{tabular}
\end{table}

As (p,c) increases, the solution tree size prevents the use of these
algorithms and other methods such as genetic algorithm come into
play. Merelo and al (e.g., \cite{Merelo2006} or \cite{Merelo2010}) or
\citet{Berghman2009} produced results for higher cases. For MM(5,8), to my
knowledge, the best upper bound found in the literature is 5.618 in
\citep{Berghman2009}, above \textit{Entropy} 5.489 in
\autoref{tab:comparealgo58}, but reached with a much lower computation
time.

\section{Optimal strategy: some general cases solved}
This section presents theoretical results already established. MM(p,1) and
MM(1,c) are obvious. MM(2,c) is solved for the general case. MM(3,c) is
solved for the pessimistic case.

\subsection{MM(p,1)}
As there is only one code $\overbrace{\text{\texttt{1\ldots1}}}^p$,
\begin{equation}
  \left.\forall p, \quad
  \begin{cases}
    E(p,1) &= 1 \\
    W(p,1) &= 1
  \end{cases}
  \right.
\end{equation}

\subsection{MM(1,c)}
\label{sec*mm1c}
If $c=1$, MM(1,c) is equivalent to the previous case and takes one
guess to solve. If $c=2$, it takes two guesses
(\comb{1}=\comb{1}(1,0) and \comb{2}=\comb{1}(0,0)-\comb{2}(1,0)).  So
$E=\frac{1+2}{2}=\frac{3}{2}$ and $W=2$.

Similarly, for $c=n$, it always takes $n$ guesses. Thus, $W=n$ and
\[E=\frac{1}{n}\times\sum_{i=1}^{n}{i}=\frac{1}{n}\times\frac{n(n+1)}{2}=\frac{n+1}{2}\]

\bigskip
Finally,
\begin{equation}
  \left.\forall c, \quad
  \begin{cases}
    E(1,c) &= \frac{c+1}{2} \\
    W(1,c) &= c
  \end{cases}
  \right.
\end{equation}

\subsection{MM(2,c)}
\citet{Chen2004} and \citet{Goddard2004} solved this model for $E$ and $W$
and obtained the following results:
\begin{equation}
  E(2,2)=2, \text{ and } \forall c \ge 3, E(2,c)=
  \left\lbrace
    \begin{aligned}
      \frac{8n^3+51n^2-74n+48}{24n^2} ,& \text{ $n$ even} \\
      \frac{8n^3+51n^2-80n+69}{24n^2} ,& \text{ $n$ odd}
    \end{aligned}
  \right.
\end{equation}

\begin{equation}
\forall c \ge 2, W(2,c)=\lfloor c/2 \rfloor+2
\label{eq:w2c}
\end{equation}

The results for $c=2,\cdots,9$, presented in
Tables~\ref{tab:expectedlength} and \ref{tab:expectedaverage}, verify these
formulas for $E$ and $W$.

As \citet{Jager2009} already noticed, all results for $W$, presented in
\citet{Goddard2004}, do not all satisfy \autoref{eq:w2c}.

\subsection{Partial MM(3,c)}
\citet{Jager2009} proved that:
\begin{equation}
\forall c \ge 5, W(3,c)=\lfloor (c-1)/3 \rfloor+4
\end{equation}
and computed $W(3,c)_{c=2,3,4}=(3, 4, 4)$.

Once again, the results for $c=2,\cdots,9$, presented in
Tables~\ref{tab:expectedlength} and \ref{tab:expectedaverage}, verify this
formula for $W$.

This article shows more results for $W$ and introduces other theoretical
inequalities in the pessimistic case.

I am unaware of any paper on $E(3,c)$.

\section{Reaching for MM(4,7)}
For higher cases, one must be\dots patient. The goal was the optimal in the
expected case, so the rest of this article will focus on more $E(p,c)$
results for certain values of $(p,c)$.

A depth-first branch and bound algorithm, similar in principle to the
depth-first backtracking algorithm used by \citet{Koyama1993} for MM(4,6),
was developed. It also explores systematically all possible combinations by
branch and goes to the lowest level\footnote{The best solution of level
  $n+1$ is always sought for each combination of level $n$.} but saves the
best value found (incumbent). As soon as the current branch reaches a
higher score, the exploration stops and a new branch is explored. A lower
bound evaluation with dynamic update is also used to stop a branch
exploration before it actually reaches a higher score: reaching a higher
predictive score is sufficient. A tight upper bound comes as a proper
incumbent and allows efficient pruning from the start. Finally, some
symmetries of the problem are also detected to avoid exploring all codes at
each levels.

Given the problem size, optimizing computation time and memory space is
crucial. Three main ideas guided this optimization:
\begin{enumerate}
\item Use as many shortcuts as possible,
\item Prune the tree as much as possible using a good upper bound from the
  start and a dynamic lower bound evaluation,
\item Use symmetry as much as possible to avoid exploring similar branches.
\end{enumerate}

\subsection{Shortcuts}
Some obvious cases encountered along the resolution can be solved right
away and can shorten the computation time. To manage the program complexity
throughout the years of this work, only simple ones were implemented.

For any set of size $k<G_p$ codes, if one of them can discriminate all of
the others, an absolute minimum is found and this guess can be played right
away. The external path is increased by $1\times1+2\times(k-1)=2k-1$ and 2
guesses are needed at most. Otherwise, if a non-possible code can do the
same (see \autoref{tab:nonpossiblecode} for such an example), the external
path is increased by one more ($1\times0+2\times k=2k$) and at most two
guesses are still needed. Each time such configurations are found, a
shortcut can be taken.

Conversely, when a code creates only one subset, nothing is gained and the
program can backtrack immediately.

There exist refinements of the first shortcut. The $k=1$ case is trivial
(just play the code!) and can be treated upstream separately. The $k=2$
case can also be treated upstream after noticing that simply playing one of
the codes followed by the second exactly represents the optimum
described. The $k=3$ case could also be treated separately but has an
impact on the number of guesses. Indeed, either one of the three codes can
segregate the other two and an optimum is found ($1\times1+2\times2=5$), or
playing each of the three codes in sequence leads to 6
($1\times1+2\times1+3\times1=6$) as playing a non-possible code segregating
all others would also lead us to ($1\times0+2\times3=6$). But in the latter
case, 3 guesses are nevertheless needed instead of 2. This $k=3$ case was
not implemented as such. This issue is addressed in the conclusion.

Another shortcut deals with an answer of (0,0) after the first guess. When
solving the possible case for MM(p,c), namely MM*(p,c), the (0,0) answers
means that none of the $k$ colors used in the first guess are part of the
code and the branch itself is therefore MM*(p,c-k). In this case, when the
program is run to obtain a value and not a solution, using previously found
results saves a lot of time. A similar scheme could be applied in the
general case. Indeed, in the same situation, but using all possible colors,
it would be the same as solving the MM(p,c-k) but with more available
colors (one in fact, as explained in \autoref{sec:symmetry}). Saved
MMe(p,c) data (e for \textit{extended}) from previous runs could also be
used in a run where the result is desired rather than the solution. This
shortcut was implemented for possible solutions and not for the general
case. Note that these values are also used for upper bounds, as explained
in the next section, and allow easy detection of discrepancies when testing
new versions of the program.

Finally, as obvious as it may seem, storing all pairwise gradings into a
table saves computing time at each node. As there is a memory/speed
trade-off here, an option to turn it off comes in handy.

\subsection{A tight upper bound from the start}
\label{sec:upperbound}
Keeping the \textit{best} value found so far allows branch pruning each
time this incumbent is exceeded. This saves a lot of unnecessary
computation and is the basis of the algorithm. But the higher the
incumbent, the more unnecessary branches are explored. A good upper bound
at the beginning accelerates the pruning process from the start.

An excellent upper bound is the optimal value for MM*(p,c). Such computing
option was introduced in the program. This version is obviously much faster
because the number of codes is reduced at each step rather than remaining
constant. The final result, while not optimal in the general
case\footnote{Note that the results are the same for 9 cases between
  \autoref{tab:expectedlength} and \autoref{tab:optipossible}.}, is already
pretty close. For example, in \autoref{tab:optipossible}, the values for
MM*(4,6) and MM*(4,7) are better than those found with the one-step-ahead
algorithms. This `possible' version of the program is fed, in turn, an
upper bound from a heuristic computation.

\subsection{A dynamic lower bound evaluation}
To be efficient, the pruning process requires, for each branch, a lower
bound evaluation beforehand and a dynamic update along its resolution. The
pruning occurs sooner as any overrun above the incumbent is not only
detected but also predicted.

A classic lower bound can be computed imagining that the remaining codes
are found using a perfect tree.  Such a tree would have a maximum branching
factor of $G_p$ at each internal node and would be perfectly
balanced. After $q$ questions\footnote{Depth is $q$-1 by definition.}, all
the nodes are leaves. For such a tree, let $L(p,q)$ denotes the external
path length and $T(p,q)$ the total number of leaves. Then,
\begin{equation}
\forall q \geq 1,\, T(p,q)=\sum_{i=1}^{q}(G_p-1)^{i-1}=\sum_{i=0}^{q-1}(G_p-1)^i
\end{equation}

One question implies 1 leaf node ($(p,0)$ immediately). Two questions
imply $G_p$ branches and $(p,0)$ for all $G_p-1$ leaves at question 2,
thus $T(p,2)=1+(G_p-1)=G_p$ is the total number of nodes and
$L(p,2)=1\times1+2\times(G_p-1)=2G_p-1$. This idea was explained in
the shortcut section.

Given a number of nodes of $M$, and if the perfect tree of M nodes can be
found in $q+1$ questions, the external path is equal to the path of $q+1$
questions minus the total number found in previous questions: 

$\forall q>0, T(p,q)<M \leq T(p,q+1) => $
\begin{equation*}
\begin{split}
L(p,q+1,M) &= \sum_{i=1}^{q}i(G_p-1)^{i-1}+(q+1)(M-T_q) \\
           &= (q+1)M-[(q+1)T_q - \sum_{i=0}^{q-1}(G_p-1)^i(i+1)] \\
           &= (q+1)M-\sum_{i=0}^{q-1}(G_p-1)^i(q+1-i-1) \\
           &= (q+1)M-\underbrace{\sum_{i=0}^{q-1}(G_p-1)^i(q-i)}_{S} \\
%           &= (q+1)M-A(p,q)
\end{split}
\end{equation*}

Let $S(p,q)$ denote the \textit{Sum} of external path lengths after q
questions in an optimal tree, \ie
\begin{equation}
\forall q\geq1,\, S(p,q)=\sum_{i=0}^{q-1}(G_p-1)^i(q-i),\, \text{ and } S(p,0)=0
\end{equation}

then, given a $p$ value, the lower bound equation becomes
\begin{equation}
T_q<M \leq T_{q+1} => L_{q+1}=(q+1)M-S_q
\label{eq:lowerbound}
\end{equation}

Both $S$ and $T$ are related by recursion equations,
\begin{equation*}
\begin{split}
\forall q\geq0, S_{q+1} &= \sum_{i=0}^{q}(G-1)^i(q+1-i) \\
                      &= 1\times(q+1)+\sum_{i=1}^{q}(G-1)^i(q+1-i) \\
                      &= (q+1)+\sum_{j=0}^{q-1}(G-1)^{j+1}(q+1-j-1) \\
                      &= (q+1)+(G-1)\sum_{j=0}^{q-1}(G-1)^j(q-j)
\end{split}
\end{equation*}

\begin{equation}
\Rightarrow S_{q+1} = (q+1)+(G-1)S_q
\end{equation}

Similarly,
\begin{equation}
T_{q+1}=T_q+(G-1)^q=1+(G-1)T_q
\label{eq:T}
\end{equation}

And the name of \textit{S} is justified by finally noticing that,
\begin{equation*}
\begin{split}
\forall q\geq0, \, S_{q+1}&=\sum_{i=0}^{q}(G-1)^i(q+1-i)\\
                         &=\sum_{i=0}^{q}(G-1)^i(q-i) + \sum_{i=0}^q(G-1)^i\\
\end{split}
\end{equation*}

\begin{equation}
\Rightarrow S_{q+1}=S_q+T_{q+1}
\label{eq:S}
\end{equation}

% When this lower bound is greater than the best kept score, the
% computation for the branch can be stopped right away. The current
% score is updated as soon as the real score is known instead of keeping
% the estimation based on an optimal branch. In this way, non optimal
% solutions are abandoned faster.

This lower bound is simple and can be computed efficiently
(\autoref{eq:lowerbound}) but it does not perform too well on big sets.

To improve its value, the computation for a node is the sum of the lower
bounds of each subset for this node. The evaluation for each subset is
replaced by its real value whenever known and the comparison with the
incumbent is done at each level instead of at the end.

There exists another way of improving the value of the lower bound.
Indeed, while a perfect tree based on the maximum branching factor $G$ can
be imagined, when trying all remaining codes against all guesses, most
often $k<G$ subsets are found. From that point on in the branch, no more
than $k$ subsets will be found. So a higher lower bound can be computed
with a branching factor of $k$ instead of $G$. The \autoref{eq:T} and
\autoref{eq:S} allow computation and caching of all values from 2 to $G$ to
improve speed.

This makes the computation highly dynamic and the score converges faster to
the real value.

\subsection{An integrated scheme to detect symmetries and use case
  equivalence}
\label{sec:symmetry}
Symmetry is a key element in the resolution.  Many articles mention
external programs like
\textit{Nauty}\footnote{\url{http://cs.anu.edu.au/~bdm/nauty/}} to reduce
the number of codes by exploiting all symmetries of the game. For speed
purposes, a trade-off had to made between refining the scheme of symmetry
detection and the time spent progressing in the resolution even though more
branches might be examined. A method based on the properties of colors at
each stage is proposed. A full comparison between the two methods has not
been made. The method described hereafter could also be a first code filter
upon which other mechanism could be implemented.

As can be understood, for the first guess, all codes but a handful need to
be tested. For example, in the seven color case, all seven one-color
combinations do not need to be played, only one suffices; the other 6 cases
are covered by symmetry, replacing the one color by any of the others. All
2-color, 3-color, 4-color, \dots $min(p,c)$-color cases are also covered in
this way.  At this stage, it is obvious that neither the colors nor the
colors order matter as everything is symmetrical. This complete symmetry is
lost once one guess is played, but some interesting properties still
remain.

At any stage, any color that has never been played before -- call it a
\textit{free} color -- is symmetrical with any other \textit{free}
color. For example, in MM(4,7) and when played after a first guess of
\comb{1234}, the three guesses \comb{1235}, \comb{1236} and \comb{1237}
form an interchangeable set of codes, as well as \{\comb{1255},
\comb{1266}, \comb{1277}\} or even \{\comb{1256}, \comb{1257}, \comb{1265},
\comb{1267}, \comb{1275}, \comb{1276}\}. Note that all exclusive \free
color codes keep the complete symmetry of the beginning.

\textit{Zero} colors also play a role. Let \textit{zero} color be any color
that is not part of the code. For example, if \comb{1223} gets an answer of
(0,0), the three combinations \comb{4511}, \comb{4512} and \comb{4531} are
all equivalent for the next guess. In fact, any \zero color can be
replaced by any other \zero color and ultimately the same one. As a
result, any exclusive \zero color guess does not provide additional
information and can be filtered out.

Another opportunity to identify \zero colors comes from the case where
$b+w=p$. All the pegs are of the right color even though some of them are
not in the proper order. Therefore all other colors are \zero colors.

A signature is assigned to each code based on the respective properties of
\free and \zero colors.  The generic\footnote{For the first guess or
  exclusive \free color codes, the order of all colors is also
  reorganized.} signature is equal to the code where any \textit{zero}
colors is replaced by the letter 'z' and a \textit{free} color is replaced,
in order of appearance, by a letter in alphabetical order.  All codes with
the same signature are \textit{case equivalent}\footnote{I did not have
  access to \citet{Neuwirth1982} who seems to be the first to have
  introduced this notion. From the description in other articles, the main
  idea of a class of codes given the history of guesses is respected.}  and
only a class representative\footnote{The first in lexical order is chosen.}
needs to be tested. Note that for a given code, its signature evolves along
a branch according to the \textit{free} and \textit{zero} colors at the
given level.  The signature mechanism is especially efficient when $c \ge
p$. Additional simplification is anticipated in the case of codes made solely
of \textit{free} and \textit{zero} colors but the required level of effort
could not be dedicated to conclude.

For example, for MM(4,7), the program starts with the well known 5
codes\footnote{These 5 codes are respectively the representatives of 7, 168
  (3+1), 126 (2+2), 1260 and 840 codes. Note that $168+126=294$. See
  \autoref{sec:sumcolors} for details} (\comb{1111}, \comb{1112},
\comb{1122}, \comb{1223} and \comb{1234}) identified by this signature
method. In the \comb{1123} branch, never more than 361 of the 2400 possible
codes are tried, 41 in the case of a (0,0) answer.

\subsection{A documented example}
The following simplified MM(3,4) output illustrates how all these pieces
work together.

\begin{verbatim}
With 3 pegs and 4 colors:
- there are 9 possible ways of grading,
- starting set has 64 possible combs,

The solver will use:
- for first level, the reduced set of 3 combs (111,112,123),
- an upper bound of 207.

Erase 111 (223>=207)
With the first set :
123 = 194
112 = 196
----- 123 (E=194,g=9)
<123,0>=1 
444 ->2 ()  E:194/S:2
<123,30>=1 
123 ->1 ()  E:194/S:3
<123,3>=2 
231 ->5 ()  E:194/S:8
<123,12>=3 
Will try 57 combs finally
112 ->9 (8)  E:195/S:17
<123,20>=9 
Will try 63 combs finally
134 ->30 (26)  E:199/S:47
<123,1>=9 
Will try 63 combs finally
244 ->29 (26)  E:202/S:76
<123,11>=12 
Will try 63 combs finally
134 ->40 (38)  E:204/S:116
<123,10>=12 
Will try 63 combs finally
244 ->41 (38)  E:207/S:157
Min reached already (E:207 or S:157 >= 207). Next one.
----- 112 (E=196,g=9)
<112,30>=1 
112 ->1 ()  E:196/S:1
<112,12>=2 
121 ->5 ()  E:196/S:6
<112,2>=5 
Symmetry: 33 versus 63
Will try 32 combs finally
232 ->15 (14)  E:197/S:21
<112,11>=8 
Symmetry: 33 versus 63
Will try 33 combs finally
123 ->24 (23)  E:198/S:45
<112,0>=8 
Symmetry: 11 versus 63
Will try 11 combs finally
334 ->26 (23)  E:201/S:71
<112,20>=9 
Symmetry: 33 versus 63
Will try 33 combs finally
123 ->30 (26)  E:205/S:101
<112,1>=14 
Symmetry: 33 versus 63
Will try 33 combs finally
233 ->47 (46)  E:206/S:148
<112,10>=17 
Symmetry: 33 versus 63
Will try 33 combs finally
134 ->58 (58)  E:206/S:206
		112 = 206
Found it in 206, starting with 112
Average = 3.22
Full path is:
[111] 112->(20)->123->(10)->111->(30)->Found
<The full solution is of no interest>
[444] 112->(0)->334->(10)->444->(30)->Found
1: 1
2: 5
3: 37
4: 21
Sols= 206
------------
\end{verbatim}

Among the 64 possible first guesses, only three class representatives are
explored at the first level: one with one color (\comb{111}), one with two
colors (\comb{112}) and one with three colors (\comb{123}). An upper bound
of 207\footnote{206+1} is already known from solving the possible case.

Immediately, \comb{111} is discarded because its lower bound of 223 is
already higher than the incumbent.

\comb{123} has a first estimated lower bound of 194. The resolution pursues
until it reaches 207. The real score at that point is only 157.

\comb{112} is pursued until the end. When the answer (0,0) is given, only
11 codes out of 63 are explored.

Note how the real branch value and the evaluation beforehand (between
brackets) are quite close for the small sets presented. It is good practice
to check the evaluation against the real value.
                                   
\section{Results}
All optimal MM(p,c) results are presented in two tables.
\autoref{tab:expectedlength} shows the expected path lengths while
\autoref{tab:expectedaverage} shows the expected averages. Note that since
MM(2,c) is solved, the first line of both tables is for validation
purposes.

Already known results (\citet{Koyama1993}, \citet{Goddard2004}) were found
together with five others unpublished to my knowledge (in bold). For
MM(4,7), $E(4,7)=11228/2401=4.676$ with a worst case of 6
guesses\footnote{The $f_i$ are (1/8/78/717/1473/124)} using \comb{1123} as
the first guess. All these results have the same worst case as the ones in
Table~1 of \citet{Jager2009} but for MM(4,6) (the famous 6 versus 5 case).

\begin{table}[ht]
  \centering
  \caption{Optimal path lengths in the expected case for MM(p,c)}
  \label{tab:expectedlength}
  \begin{tabular}{|c|n{3}{0}|n{4}{0}|*{6}{n{5}{0}|}} \hline
\diagbox{p}{c} & \ccell{2} & \ccell{3} & \ccell{4} & \ccell{5} & \ccell{6} & \ccell{7} & \ccell{8}     & \ccell{9}     \\ \hline
2 & 8   & 21         & 45         & 81   & 132  & 198         & 284        & 388        \\ \hline
3 & 18  & 73         & 206        & 451  & 854  & 1474        & \myb{2359} & \myb{3596} \\ \hline
4 & 44  & 246        & 905        & 2463 & 5625 & \myb{11228} &            &            \\ \hline
5 & 97  & 816        & \myb{3954} &      &      &             &            &            \\ \hline
6 & 224 & \myb{2649} &            &      &      &             &            &            \\ \hline
7 & 496 &            &            &      &      &             &            &            \\ \hline
  \end{tabular}
\end{table}

\begin{table}
  \centering
  \caption{Optimal results in the expected case ($E(p,c$))}
    \label{tab:expectedaverage}
    \begin{tabular}{|c|*{8}{n{1}{3}|}} \hline
\diagbox{p}{c} & \ccell{2} & \ccell{3} & \ccell{4} & \ccell{5} & \ccell{6} & \ccell{7} & \ccell{8}     & \ccell{9}     \\ \hline
2 & 2.000 & 2.333       & 2.813       & 3.240 & 3.667 & 4.041       & 4.438       & 4.790       \\ \hline
3 & 2.250 & 2.704       & 3.219       & 3.608 & 3.954 & 4.297       & \myb{4.607} & \myb{4.933} \\ \hline
4 & 2.750 & 3.037       & 3.535       & 3.941 & 4.340 & \myb{4.676} &             &             \\ \hline
5 & 3.031 & 3.358       & \myb{3.861} &       &       &             &             &             \\ \hline
6 & 3.500 & \myb{3.634} &             &       &       &             &             &             \\ \hline
7 & 3.875 &             &             &       &       &             &             &             \\ \hline
    \end{tabular}
\end{table}

\autoref{tab:optipossible} shows the expected path lengths for all optimal
results in the possible case. This table contains 8 more upper bonds (in
bold). These values may be helpful when comparing with other algorithms or
even trying to solve the optimal case.  For MM*(5,6), only one code
requires 7 steps. I therefore believe that $W(5,6) \leq 6$. MM*(5,8) is yet
to be found to improve the upper bound of 5.489 found in
\autoref{sec:heuristics}.

\begin{table}
  \centering
  \caption{Optimal path lengths in the expected case for MM*(p,c)}
  \label{tab:optipossible}
  \begin{tabular}{|c|n{3}{0}|n{4}{0}|*{6}{n{5}{0}|}} \hline
\diagbox{p}{c} & \ccell{2} & \ccell{3} & \ccell{4} & \ccell{5} & \ccell{6} & \ccell{7} & \ccell{8}     & \ccell{9}     \\ \hline
2 & 8   & 21         & 45          & 81          & 134         & 205         & 299         & 417         \\ \hline
3 & 18  & 73         & 206         & 455         & 864         & 1503        & 2439        & 3749        \\ \hline
4 & 44  & 247        & 908         & 2476        & 5660        & 11362       & \myb{20838} & \myb{35426} \\ \hline
5 & 97  & 824        & 3982        & \myb{13572} & \myb{36920} & \myb{86270} &             &             \\ \hline
6 & 225 & 2671       & \myb{17416} & \myb{74140} &             &             &             &             \\ \hline
7 & 505 & \myb{8817} &             &             &             &             &             &             \\ \hline
  \end{tabular}
\end{table}

Finally, \autoref{tab:expectedlengthzero} presents the results from solving
MMe(p,c) (solving MM(p,c) with more than $c$ colors) at guess 2. Most of
these results are extracted from runs in the case where a first guess
received a (0,0) answer. These results could be used to save time when
solving higher cases. They were corrected\footnote{The computation of $L$
  does not start at guess one but at guess 2, $\sum_{i=1}^{w}{(i+1)
    f_i}=\sum_{i=1}^{w}{if_i}+\sum_{i=1}^{w}{f_i}=L+N$} for direct
comparison with \autoref{tab:expectedlength}. Results are the same but for
the cells in gray. For these `gray' cases, a \textit{zero} color is used
among the guesses, most often in the first one (second in the global
resolution) when for the others only \textit{free} colors are used. It
seems to indicate that MM(p,c)=MMe(p,c) whenever $c \geq p-1$, \ie the
fewer the colors compared to the number of pegs, the sooner a \textit{zero}
color is needed to discriminate the remaining codes. This conjecture has to
be assessed on more examples and confirmed theoretically.

\newcommand{\myg}[1]{{\cellcolor[gray]{.8}\color{black}}#1}
\begin{table}[ht]
  \centering
  \caption{Optimal path lengths in the expected case for MMe(p,c)}
  \label{tab:expectedlengthzero}
  \begin{tabular}{|c|n{1}{0}|n{3}{0}|n{4}{0}|*{5}{n{5}{0}|}} \hline
\diagbox{p}{c} & \ccell{1} & \ccell{2} & \ccell{3} & \ccell{4} & \ccell{5} & \ccell{6} & \ccell{7}     & \ccell{8}    \\ \hline
2 & 1 & 8         & 21         & 45    & 81   & 132 & 198  & 284  \\ \hline
3 & 1 & 18        & 73         & 206   & 451  & 854 & 1474 & 2359 \\ \hline
4 & 1 & \myg{40}  & 246        & 905   & 2463 &     &      &      \\ \hline
5 & 1 & \myg{91}  & \myg{815}  &       &      &     &      &      \\ \hline
6 & 1 & \myg{189} & \myg{2646} &       &      &     &      &      \\ \hline
7 & 1 & \myg{412} &            &       &      &     &      &      \\ \hline
  \end{tabular}
\end{table}

\section{Conclusion}
An optimal MM(4,7) strategy was found along with other optimal strategies
in the expected case. Additional tight upper bounds (optimal in the
possible case) for other cases are also presented as well as an upper bound
for MM(5,8).

The signature scheme is an efficient way of only testing a class
representative and reducing the number of codes tried at each step. The
dynamic lower bound mechanism also gives good results. The MM*(p,c) upper
bound is a good starting value.

After many years of pursuing an optimal MM(4,7) strategy, the program in
its present form has reached its limits in terms of speed and memory
space. Perl was used to easily test and implement new ideas over the years
while managing the complexity of the program. Further results require a
faster and memory-optimized language.

Following are a few ideas that have not been implemented but should enhance
the resolution.

The code signature when dealing exclusively with \textit{free} and
\textit{zero} colors should be studied and may lead to less
representatives.

A generic program for both computing and finding the optimal is not the
best solution. To illustrate, let's return to the case where 3 codes are
left. We saw that playing each code one by one would lead to the same $L$,
but with more guesses, than a true discriminating code as in
\autoref{tab:nonpossiblecode}. A first phase to find the optimal $L$
followed by a second one to find the solution itself, with a possible lower
tree depth, would be globally faster. In the second phase, the first guess
is assumed to be the same and the minimal external length is already known.

This would allow the implementation of the (0,0) first answer for the
generic case and not only for the possible case. This idea by itself can
lead to some theoretical work that would further help understand the
mechanism of \textit{zero} and \textit{free} colors.

Following the same idea, I am certain that more results could also be
stored and reused in higher cases. This is where external programs could
play an important role to detect such cases and cut the tree by solution
blocks. The case of $b+w=p$ at the first step, while marginal, is such an
example. Or other cases where $k$ never-used-before colors are tried and
obtain a (0,0) answer even at a second or further guess. The MMe results
could then also be applied.

As computer memory grows, classic computer-algorithm optimizations
could also be implemented. More intermediate computations could be
saved and used along the way and a two-step ahead mechanism could be
programmed to go for the best guess first (using the fact that playing
\comb{a} then \comb{b} is equivalent to playing \comb{b} and \comb{a}
for the rest of the branch). End games could also be introduced for
that matter. This overhead is acceptable for bigger size problems.

Such optimizations would lead to a three-step resolution.  A good starting
upper bound would be computed through any heuristic algorithm. The first
step would use this value to run the possible version with all possible
optimizations to find a tighter upper bound. The second step, with all
codes, all shortcuts and all optimizations would only search for the
optimal $L$, no solution would be recorded\footnote{By itself, this speeds
  up the process.} but the first two guesses leading to this
value. Finally, once the optimal value is found, a last pass would find a
solution by exploring a small search space using the saved guesses. This
last one could eventually focus on a low worst case.

I hope these results will help current researchers of this field but also
give ideas to newcomers, as \citet{Rosu97} did for me at the start, several
years ago.

\bibliographystyle{plainnat}
\bibliography{biblio}

\end{document}